# Direct imaging of stress tensor around single dislocation in diamond

Takeyuki. Tsuji[1], Shunta Harada[2], Tokuyuki Teraji [3*]

[1] International Center for Young Scientists, National Institute for Materials Science, 1-1 Namiki, Tsukuba, Ibaraki 305-0044, Japan
[2] Institute of Materials and Systems for Sustainability, Nagoya University, Furo-cho, Nagoya 464-8601, Japan
[3] Research Center for Electronic and Optical Materials, National Institute for Materials Science, 1-1 Namiki, Tsukuba, Ibaraki 305-0044, Japan

**ABSTRACT**.
Dislocations are fundamental crystal defects whose stress fields govern a wide range of material properties. The analytical form of the stress tensor around single dislocation was established by elasticity theory more than 80 years ago and has provided a theoretical basis for evaluating essential characteristics of dislocations. However, direct experimental verification has long remained out of reach because it has been difficult to measure the components of the stress tensor with conventional methods. Here, we present the experimental visualization of the stress tensor around single dislocation in diamond. Using quantum sensors based on nitrogen-vacancy (NV) centers, we mapped the shear components ($\sigma_{xy}$, $\sigma_{yz}$, $\sigma_{zx}$) together with the trace of the stress tensor ($\sigma_{xx} + \sigma_{yy} + \sigma_{zz}$) around single 45° dislocation. The observed distributions exhibited good agreement with predictions from elasticity theory, thus providing experimental validation of this theoretical framework.

Dislocations are line defects broadly observed in crystalline solids [1], including structural materials [2,3], semiconductor [4,5] and ceramics [6]. The stress field caused by dislocations play a central role in plastic deformation and strongly affect not only mechanical properties but also electrical [4,7] and magnetic properties [8]. Since the concept of dislocations was first introduced by Taylor [9], Orowan [10], and Polanyi [11] in 1934, research on dislocations has remained a classical topic in materials science.

In principle, stress in a crystal is described by the stress tensor $\sigma_{ij}$ (i,j = x,y,z), a second-rank tensor with nine components. Owing to the symmetry condition $\sigma_{ij} = \sigma_{ji}$, six of these components are generally independent. The analytical form of the stress tensor around a dislocation was derived from elasticity theory before 1942 [12], providing a theoretical basis for evaluating essential characteristics of dislocations, such as their elastic energy, interaction forces, and growth directions [12]. Consequently, the stress tensor around dislocations have served as a cornerstone for understanding the mechanical properties of dislocations for more than 80 years [12].

Despite its importance, the stress tensor around a dislocation predicted by elastic theory [12] has not been experimentally verified because it has been difficult to resolve the components of stress tensor with conventional measurement methods. Raman spectroscopy evaluates stress through Raman peak shifts caused by changes in lattice vibration, typically assuming only hydrostatic pressure is present ($\sigma_{xx} = \sigma_{yy} = \sigma_{zz}$, $\sigma_{xy} = \sigma_{yz} = \sigma_{xz} = 0$). With confocal optics, Raman spectroscopy enables three-dimensional stress mapping [13]. Birefringence microscopy detects strain-induced changes in the refractive index by using transmitted light and is sensitive to in-plane shear stress [14]. X-ray topography detects strain as signal intensity variations in X-ray diffraction, but cannot distinguish individual stress-tensor components [15]. Transmission electron microscopy (TEM) provides high-resolution images of dislocation structures [16], while electron backscatter diffraction (EBSD) detects local stress through distortions in Kikuchi patterns, with sensitivity mainly to shear stress near the surface because of limited electron penetration depth [17,18]. None of these techniques, however, is capable of reconstructing the components of the stress tensor.

In contrast, nitrogen-vacancy (NV) centers in diamond provide a quantum sensing platform capable of resolving stress-tensor components. The measurement principle [19–21] is as follows. An NV center is a point defect consisting of a substitutional nitrogen atom adjacent to a vacancy and has a spin-1 electronic ground state. When stress is applied to an NV center, the two resonance frequencies of NV center shift through a spin-mechanical interaction between its electron spin and the stress. Due to variations in the arrangement of the nitrogen and vacancy, NV centers can be arranged in four crystallographic directions within the diamond, corresponding to $i = 1, 2, 3, 4$ in Fig. 1 (a). Thus, a given stress field produces $4 \times 2 = 8$ distinct resonance shifts in total. By using optical detected magnetic resonance (ODMR) to measure these shifts [22], we can reconstruct the components of the stress tensor ($\sigma_{xy}$, $\sigma_{yz}$, $\sigma_{zx}$, $\sigma_{xx}+\sigma_{yy}+\sigma_{zz}$) with submicron spatial resolution. [19].

*Corresponding author: TERAJI.Tokuyuki@nims.go.jp

In this study, we employed NV centers in diamond to measure the stress tensor around a single 45° dislocation. The measured stress fields were directly compared with the theoretical predictions of elasticity theory, providing experimental validation of the theoretical predictions from elasticity theory.

In order to create NV centers around a dislocation, we synthesized $^{12}$C-enriched nitrogen-doped chemical vapor deposition (CVD) diamond film approximately 20-μm thick on a 500-μm-thick (001) diamond substrate by using microwave plasma chemical vapor deposition. The density of the NV centers in the film was estimated to be approximately 0.05 ppm. (The detailed growth conditions are described in the supplemental material.)

We performed X-ray topography to evaluate the Burgers vector of a dislocation, **b**, in the film. Metal that was approximately 500-μm thick was placed in the path of the X-rays to prevent the substrate from being irradiated and thereby the image would be of the film (see Fig. S1 in the supplement material).

Here, we briefly explain the experimental procedure used to image stress-tensor components around the dislocation. The NV center's electronic ground-state Hamiltonian in the presence of stress and a static magnetic field is [23]

$$H = (D + M_{Z_i})S_{Z_i}^2 + \gamma \vec{B} \cdot \vec{S_i}$$
$$-M_{X_i}(S_{X_i}^2 - S_{Y_i}^2) + M_{Y_i}(S_{X_i}S_{Y_i} + S_{Y_i}S_{X_i}), \quad (1)$$

where $D \approx 2.87$ GHz is the temperature-dependent zero-field splitting parameter, $\gamma=28.03$ GHz/T is the NV gyromagnetic ratio, $\vec{S_i} = (S_{X_i}, S_{Y_i}, S_{Z_i})$ are the spin-1 operators, $\vec{B}$ is the applied magnetic field, and $\vec{M} = (M_{X_i}, M_{Y_i}, M_{Z_i})$ is the spin-stress interaction, (X, Y, Z) represent the coordinate system for the particular NV orientation as shown in Fig.S2 of the supplement material. The resonance frequencies of each orientation of the NV center are given by

$$(f_\pm)_i = D + M_{Z_i} \pm \sqrt{(\gamma B_{Z_i})^2 + (M_{X_i})^2 + (M_{Y_i})^2} \quad (2)$$

where $B_{Z_i}$ ($i = 1,2,3,4$) is the magnetic field applied parallel to each NV center's axis shown in Fig. 1 (a) [19]. Z component of the spin-stress interaction, $M_{Z_i}$, is described

$$M_{Z_i} = a_1 (\sigma_{xx} + \sigma_{yy} + \sigma_{zz})$$
$$+2a_2(p_i\sigma_{xy} - q_i\sigma_{xz} - p_iq_i\sigma_{yz}) \quad (3)$$

where the stress susceptibility parameters are $a_1= 4.86$

*Corresponding author: TERAJI.Tokuyuki@nims.go.jp

and $a_2= -3.7$ (MHz/GPa) [19,21] and $\sigma_{xx}$, $\sigma_{yy}$, $\sigma_{zz}$, $\sigma_{xy}$, $\sigma_{yz}$ and $\sigma_{zx}$ are stress-tensor components with respect to the diamond unit coordinate system (x=[100], y=[010], z=[001]). As the orientation of the NV centers are represented using the subscript $i = 1, 2, 3, 4$ in Fig. 1(a), we have $(p_1, p_2, p_3, q_4)=(+1, -1, -1,+1)$ and $(q_1, q_2, q_3, q_4)=(-1, +1, -1,+1)$. First, a magnetic field of approximately 7.2 mT was applied to enable the eight resonance frequencies of the NV centers to be observed in the ODMR spectrum. Each resonance frequency $(f_\pm)_i$ ($i = 1,2,3,4$) was obtained by using a Ramsey sequence to improve the measurement accuracy. (The Ramsey sequence is described in the supplemental material, as are the details of the experiment setup and results for Ramsey fringes.) We defined the average of two resonant frequencies $S_i$ as

$$S_i = \frac{f_{+i} + f_{-i}}{2}. \quad (4)$$

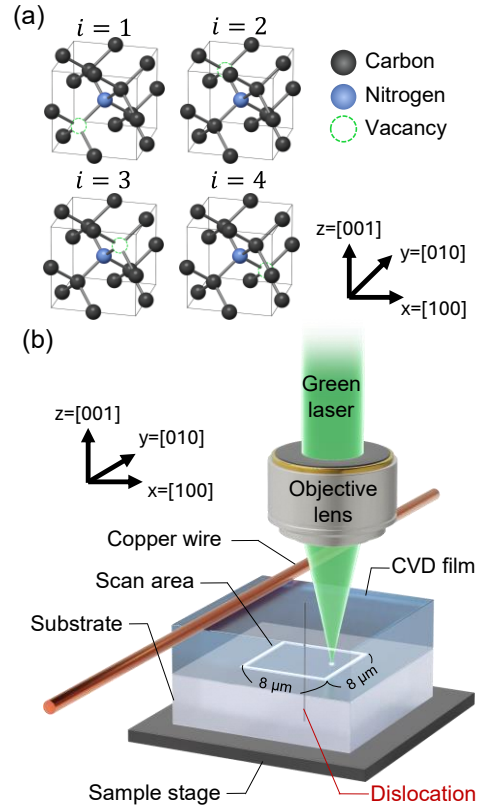

FIG 1. (a) NV centers in diamond along four different arrangements. (b) Schematic diagram of the experimental setup: CVD diamond film including nitrogen-vacancy (NV) centers was illuminated by a green laser focused by an objective lens. A copper wire was placed near the dislocation to irradiate the microwaves. The diamond was mounted on a sample stage.

We determined the three shear stress components ($\sigma_{xy}$, $\sigma_{yz}$, $\sigma_{zx}$) and the trace of the stress tensor ($\sigma_{xx} + \sigma_{yy} + \sigma_{zz}$) by using

$$\sigma_{xy} = \frac{S_1 - S_2 - S_3 + S_4}{8a_2}, \quad (5)$$

$$\sigma_{yz} = \frac{S_1 + S_2 - S_3 - S_4}{8a_2}, \quad (6)$$

$$\sigma_{zx} = \frac{S_1 - S_2 + S_3 - S_4}{8a_2}, \quad (7)$$

$$\sigma_{xx} + \sigma_{yy} + \sigma_{zz} = \frac{\frac{S_1 + S_2 + S_3 + S_4}{4} - D}{a_1}, \quad (8)$$

Compressive stress and tensile stress were defined as positive and negative, respectively. The resonance frequencies of the NV centers were detected using a confocal microscope, as shown in Fig. 1 (b). By moving the sample stage and repeating the above procedure, we obtained a mapping of the stress tensor around the dislocation.

Figures 2 (a)-(f) show the X-ray topography images of the CVD diamond film. The diffraction vectors **g** of these images are **g**=[404], [$\bar{4}$04], [$\bar{1}$13], [113], [0$\bar{4}$4], and [044], respectively. The dislocation is visible in the image when **b·g**≠0, whereas it disappears when **b·g**=0. The defect indicated by the black circles in (a)-(f) were located at the same position within the film. Thus, images (a)-(f) are of the same area in the diamond film. The image of the defect indicated by the red circles in (b)-(f) disappeared when **g** was [404] (Fig. 2 (a)), indicating that this defect was a dislocation with a Burgers vector of **b**=[$\bar{1}$01].

Figure 3(a) shows a fluorescence image of a cross section (xz plane) of the CVD diamond film that was obtained with the confocal microscope (Fig.1 (b)). Since the CVD film contained NV centers, high fluorescence intensity was observed in this region. Figure 3 (b) shows fluorescence images in the xy plane. The dislocation was located near the center of this image. To identify the position of the dislocation using the confocal microscope setup, we used X-ray topography images, birefringence microscopy images, and optical microscopy images. (The identification procedure is detailed in the supplementary material.) Figure 3 (c) shows the continuous-wave ODMR spectrum measured at position P indicated in Fig. 3 (b). The spectrum clearly shows the eight resonance frequencies $(f_{\pm})_i$ ($i = 1,2,3,4$) of the NV centers. We evaluated the dislocation line vector by imaging the resonance frequency of $f_{+3}$ in the CVD diamond film. Figure. 3(d) shows an image of $f_{+3}$ in the xy plane of the film. The $f_{+3}$ changed by approximately 0.6 MHz due to the stress field caused by the dislocation near its center. Fig. 3(e) shows an image of $f_{+3}$ along a cross-

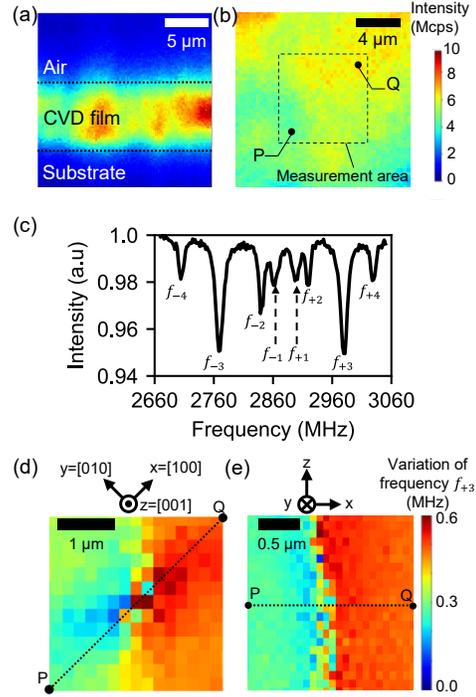

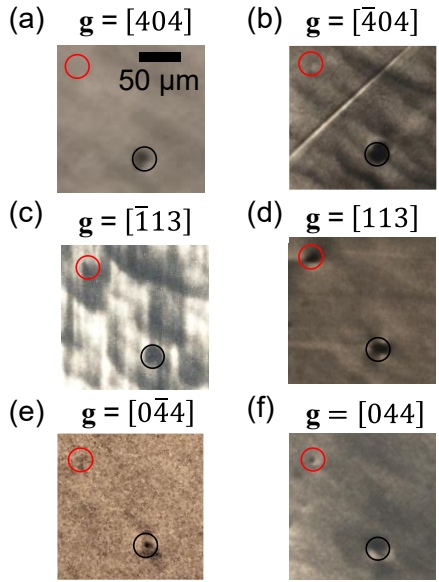

FIG 2 (a)-(f) X-ray topography images of CVD diamond film for various diffraction vectors **g**.

FIG 3. (a) Fluorescence image of the cross-section (xz plane) of the CVD film using the confocal microscope. (b) Fluorescence image of the xy plane within the CVD diamond film. The dislocation was located near the center of this image. Imaging the stress-tensor components was performed in the 8 μm × 8 μm region enclosed by the dotted line. (c) ODMR spectrum measured at positions P in Fig. 3 (b). (d) Mapping of $f_{+3}$ in the xy plane of the CVD diamond film. (e) Mapping of $f_{+3}$ in the cross-section of CVD diamond film that contains the line segment P-Q indicated in Fig. 3(d).

*Corresponding author: TERAJI.Tokuyuki@nims.go.jp

section of the film that contains the line segment P-Q indicated in Fig. 3(d). The significant change in $f_{+3}$ observed near the center of the dislocation was aligned parallel to the z = [001] direction. This indicates that the dislocation line vector **l** was [001] direction. As mentioned earlier, the X-ray topography results indicate that the Burgers vector of this dislocation **b** was [$\bar{1}$01]. Thus, this dislocation was a 45° dislocation because the angle between **b** and **l** was 45°. Previous study using X-ray topography [24] have reported that 45° dislocations with **b**=[10$\bar{1}$] dominantly exist in (001) CVD diamond films. N. Fujita *et al.* [25] performed *ab initio* modelling showing that 45° dislocations with a [001] dislocation line vector are far more stable than pure edge and crew dislocations, making our result consistent with theirs.

Finally, we divided the 8 μm × 8 μm area around the dislocation into a grid of 10 × 10 pixels in the xy plane. Then, we measured the stress tensor in each pixel using the procedure described above. Figure 4 (a) shows the stress-tensor components ($\sigma_{xx}+\sigma_{yy}+\sigma_{zz}$, $\sigma_{xy}$, $\sigma_{yz}$, $\sigma_{zx}$) around the dislocation. Here, to focus on the stress caused by the dislocations, we subtracted the stress values at a position far from the dislocation as background. (Details are in the supplemental material.) The variations in $\sigma_{xx}+\sigma_{yy}+\sigma_{zz}$, $\sigma_{xy}$, $\sigma_{yz}$, and $\sigma_{zx}$ were approximately ± 0.03, ± 0.01, ± 0.01, and ± 0.01 GPa, respectively. The measurement errors of $\sigma_{xx}+\sigma_{yy}+\sigma_{zz}$, $\sigma_{xy}$, $\sigma_{yz}$, and $\sigma_{zx}$ were approximately ten times lower than the variations. (Images of the measurement error around the dislocation are shown in the supplemental material.)

Here, we examine the degree of agreement between the experimental result and elasticity theory [12]. In elastic materials, two fundamental types of dislocations exist: edge dislocations, where the angle between the Burgers vector and the dislocation line vector is 90°, and screw dislocations, where the angle is 0°. Elasticity theory has assumed that the stress tensor of a 45°dislocation can be represented as a linear combination of the stress tensors of these two dislocation types [12]. The stress-tensor components of single 45 ° dislocation in diamond have been expressed using elastic theory [12] as

$$\sigma_{xx} + \sigma_{yy} + \sigma_{zz} = -\frac{\mu b}{2\pi(1-\nu)} \frac{x^2 y + x^3 + \nu y(x^2+y^2)}{(x^2+y^2)^2} \cos(45°) \quad (9)$$

$$\sigma_{xy} = -\frac{\mu b}{2\pi(1-\nu)} \frac{x(x^2-y^2)}{(x^2+y^2)^2}\cos(45°) \quad (10)$$

$$\sigma_{yz} = -\frac{\mu b}{2\pi} \frac{y}{x^2+y^2}\cos(45°) \quad (11)$$

$$\sigma_{zx} = \frac{\mu b}{2\pi} \frac{x}{x^2+y^2}\cos(45°) \quad (12)$$

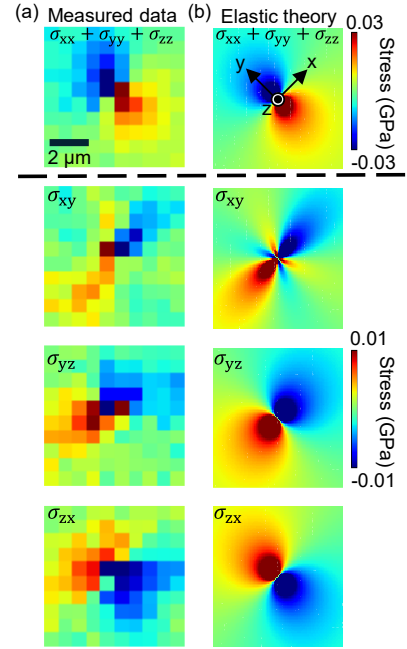

Fig. 4 (a) Experimental data for imaging the stress tensor components ($\sigma_{xx}+\sigma_{yy}+\sigma_{zz}$, $\sigma_{xy}$, $\sigma_{yz}$, $\sigma_{zx}$) around the dislocation. (b) The analytical results of the stress tensor components around a 45°dislocation with Burgers vector of [10$\bar{1}$] by using elasticity theory.

where x = [100] and y = [010] are coordinates centered on the dislocation, $\mu$=553 (GPa) is the Lamé constant of diamond, $\nu$=0.070 is Poisson's ratio of diamond [26], and $b=\left|\frac{a}{2}[10\bar{1}]\right| = \frac{\sqrt{2}}{2}a$ ($a$=0.356 nm) is the magnitude of the Burgers vector [25,27]. (The detailed derivation is provided in the supplemental material.) Figure 4 (b) shows the theoretically calculated stress tensor of single 45° dislocation. The magnitude and the distribution of the stress-tensor components obtained from the NV center (Fig. 4 (a)) were comparable to that derived from elastic theory. In addition, we fitted the experimental data (Fig. 4 (a)) to equations (9)-(12) to evaluate how well they agreed with elasticity theory. Specifically, we obtained the fitting parameters of the Lamé constant $\mu$ and Poisson's ratio $\nu$ and examined whether these parameters were consistent with values calculated from elasticity and moduli data in previous studies. Here, we used $\mu$ and $\nu$ as fitting parameters and set the magnitude of the Burgers vector $b=\left|\frac{a}{2}[10\bar{1}]\right| = \frac{\sqrt{2}}{2}a$ ($a$=0.356 nm) to be a constant value. As a result, we obtained $\mu$=504 ± 17 and $\nu$=0.036 ± 0.033. (Details of the fitting are described in the supplemental material.) Note that Poisson's ratio for diamond is approximately 0.07, making 1 - $\nu$ ≈ 1 in equations (9) and (10). This results in a large fitting error of 0.033 compared with the actual $\nu$ of 0.036. The Lamé constant $\mu$ is estimated to be 477 [28], 475 [29], 533 [30], 523 [26] (Gpa) and Poisson's ratio $\nu$ is 0.10 [28], 0.10 [29], 0.07 [30], 0.057 [26] based on data reported in previous studies. (The derivations of the Lamé constant and Poisson's ratio based on the data from the previous studies are described in the supplemental material.) These values are consistent with our results: $\mu$=504 ± 17 and $\nu$=0.036 ± 0.033. Therefore, we consider that this study experimentally verified the stress tensor around a dislocation derived from elasticity theory.

*Corresponding author: TERAJI.Tokuyuki@nims.go.jp

The measurement technique of stress tensor performed in this study was made possible primarily by the confocal microscopy in combination with the Ramsey sequence, which enables high-precision measurement of the resonant frequencies. Previous studies have measured residual stress in CVD diamond films by using CCD camera-based imaging and a continuous-wave ODMR scheme, which have been sensitive to stress near the diamond surface, and have achieved stress measurement error of about 0.1–1.0 MPa [19,21]. Our earlier work [31] utilized confocal microscopy with a continuous-wave ODMR scheme and realized three-dimensional imaging of the stress tensor in the CVD diamond. However, the stress measurement errors of our earlier work [36] were approximately 10 MPa, making it difficult to measure the stress tensor near dislocations where the stress variation is about 10 MPa, as shown in Fig. 4. In this study, we introduced the Ramsey sequence to improve the readout sensitivity to the resonant frequencies, and we succeeded in reaching stress measurement errors of approximately 1.0 MPa. This advancement enabled direct three-dimensional imaging of the stress tensor around single dislocation.

Effective strategies for suppressing dislocation-mediated plasticity—such as solid-solution strengthening [32], precipitation strengthening [33], dislocation interaction strengthening [34], and grain refinement [35]—have been theoretically described on the basis of the stress tensor. Consequently, techniques enabling three-dimensional precise imaging of the stress tensor have been sought. Thus, we believe that our technique to measure stress tensor becomes a powerful tool to probe plasticity at its origin, paving the way for the informed design for high-strength materials.

In conclusion, we performed direct imaging of the stress tensor around single 45° dislocation in diamond using NV center. We found that the variation in $\sigma_{xx}+\sigma_{yy}+\sigma_{zz}$, $\sigma_{xy}$, $\sigma_{yz}$, and $\sigma_{zx}$ was approximately ± 0.03, ± 0.01, ± 0.01, and ± 0.01 GPa, respectively. In addition, the distribution of stress exhibited a point-symmetric shape centered around the dislocation. The measurement results were consistent with those derived from elasticity theory. Therefore, this study offers a experimental confirmation of the stress tensor around single dislocation, validating the predictions of elasticity theory.

*Corresponding author: TERAJI.Tokuyuki@nims.go.jp

ACKNOWLEDGMENTS
We gratefully acknowledge constructive discussions provided by Dr. Junichi Inoue and Dr. Takahito Ohmura of NIMS. This work was partially supported by MEXT Q-LEAP (JPMXS0118068379), JST Moonshot R&D (JPMJMS2062), CSTI SIP "Promoting the application of advanced quantum technology platforms to social issues,", JST ASPIRE (JPMJAP24C1), and JSPS KAKENHI grants (Nos. 24H00406, 24K22963).

*Corresponding author: TERAJI.Tokuyuki@nims.go.jp

# Supplemental Materials for "Direct imaging of stress tensor around single dislocation in diamond"

Takeyuki. Tsuji[1], Shunta Harada[2], Tokuyuki Teraji [3*]

[1] *International Center for Young Scientists, National Institute for Materials Science, 1-1 Namiki, Tsukuba, Ibaraki 305-0044, Japan*

[2] *Institute of Materials and Systems for Sustainability, Nagoya University, Furo-cho, Nagoya 464-8601, Japan*

[3] *Research Center for Electronic and Optical Materials, National Institute for Materials Science, 1-1 Namiki, Tsukuba, Ibaraki 305-0044, Japan*

## 1. Details of the experiment method

### A. Diamond sample

A high-pressure and high-temperature (HPHT) type-IIa (100) single crystal diamond with a thickness of 500 μm (electronic grade, supplied by Element Six, Ltd.) was used as a substrate. The top surface of the substrate was mechanically polished along the [110] direction under a fine polishing condition. The CVD growth condition was as follows: 110 Torr reaction pressure, 1.4 kW microwave power, 10% $^{12}$C purified methane concentration ratio (flow rate ratio of $CH_4$ to the total gas flow), 10% nitrogen concentration ratio (flow rate ratio of $N_2$ to the total gas flow), 2% oxygen concentration (flow rate ratio of $O_2$ to the total gas flow), and 1020–1090°C substrate temperature. As a result, a 20 μm-thick CVD homoepitaxial diamond was obtained. The nitrogen density in the CVD diamond film was approximately 5 ppm.

### B. X-ray topography

We carried out X-ray topography (XRT) to identify the Burgers vector of a dislocation in the CVD diamond film grown on type-IIa (100) single crystal diamond. Grazing incident reflection synchrotron XRT was performed at the BL8S2 beamline of the Aichi Synchrotron Radiation Center, Japan, by using a monochromatic x-ray beam of 14.0 keV. A molybdenum plate with a thickness of approximately 500 μm was placed in the path of the X-rays to block irradiation of the substrate, allowing only the X-rays reflected from the CVD diamond films to be detected, as shown in Fig. S1. Topographic images were recorded on nuclear emulsion plates and digitally captured using an optical microscope with transmission illumination.

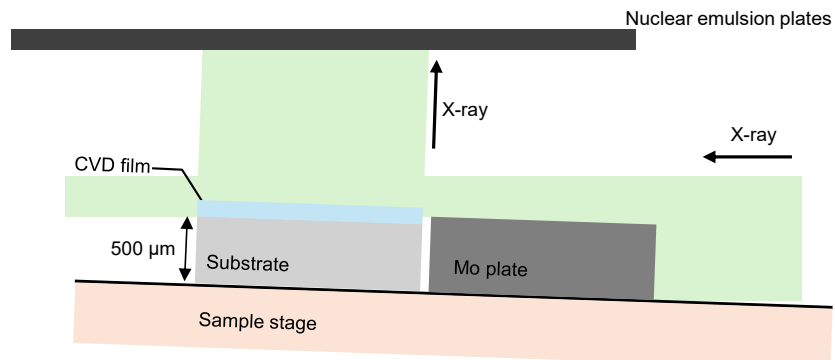

Fig. S1 Schematic setup for X-ray topography.

### C. Details of the confocal microscopy setup

The fluorescence from the NV centers was detected using a confocal microscope. The configuration of the diamond sample, objective lens (PLAPON60XO, Evident Scientific, K.K.) with a numerical aperture of 1.42, and copper wire during the measurement are shown in Fig. 1(b) of the main text. The diamond sample was mounted on a sample stage that could be scanned in the x, y, and z directions. We used a 514-nm laser (STRADUS514-60, Vortran Laser Technology, Inc.) with polarization aligned in the [010] direction. The CVD diamond film was irradiated with a laser of approximately 3 mW. The fluorescence from the NV centers was detected using an avalanche photo diode (COUNT, Single Photon Counting Module, Laser Components, GmbH) Considering the NV center density of 0.05 ppm and the detection volume of 0.5 $\mu m^3$ in the confocal microscope used in this study, it was estimated that there were approximately $10^4$ NV centers at one measurement point. The copper wire with a diameter of 20 μm was used for microwave irradiation. We used a cylindrical Samarium-cobalt magnet, Φ32× 7 (mm), with surface flux magnetic density of 210 mT to apply the magnetic field. The magnet was placed at a distance of approximately 5 cm from the CVD diamond film. A magnetic field of approximately 7.2 mT was applied to the CVD diamond.

### D. Derivation of equations (9)-(12) in the main text.

The NV center's electronic-ground-state Hamiltonian in the presence of stress and a static magnetic field [1] is

$$H = (D + M_{Z_i})S_{Z_i}^2 + \gamma \vec{B} \cdot \vec{S_i} - M_{X_i}(S_{X_i}^2 - S_{Y_i}^2) + M_{Y_i}(S_{X_i}S_{Y_i} + S_{Y_i}S_{X_i}) \quad (1-1),$$

where $D \approx 2.87$ GHz is the temperature-dependent zero-field splitting parameter, $\gamma = 28.03$ GHz/T is the NV gyromagnetic ratio, $\vec{S_i} = (S_{X_i}, S_{Y_i}, S_{Z_i})$ are the spin-1 operators, $\vec{B}$ is the applied magnetic field, and $\vec{M} = (M_{X_i}, M_{Y_i}, M_{Z_i})$ is the spin-stress interaction, ($X_i$, $Y_i$, $Z_i$) represent the coordinate system for the particular NV orientation as shown in Fig S2.(a). The relation between $\vec{M}$ and the stress tensor components ($\sigma_{xx}$,

$\sigma_{yy}$, $\sigma_{zz}$, $\sigma_{xy}$, $\sigma_{yz}$, $\sigma_{zx}$) with respect to the diamond unit coordinate system (x=[100], y=[010], z=[001]) is

$$M_{X_i} = b\left(-\sigma_{xx} + \sigma_{yy} + 2\sigma_{zz}\right) + c\left(2p_i\sigma_{xy} + q_i\sigma_{xz} + p_iq_i\sigma_{yz}\right) \quad (1-2),$$

$$M_{Y_i} = \sqrt{3}b\left(\sigma_{xx} - \sigma_{yy}\right) + \sqrt{3}c\left(q_i\sigma_{xz} - p_iq_i\sigma_{yz}\right) \quad (1-3),$$

$$M_{Z_i} = a_1\left(\sigma_{xx} + \sigma_{yy} + \sigma_{zz}\right) + 2a_2\left(p_i\sigma_{xy} - q_i\sigma_{xz} - p_iq_i\sigma_{yz}\right) \quad (1-4),$$

where the stress susceptibility parameters are $a_1$= 4.86, $a_2$= -3.7, $2b$ = -2.3, $2c$ = 3.5 (MHz/GPa) [2]. As the orientation of the NV centers are represented using the subscript $i$ ($i = 1,2,3,4$) in Fig. 1(a) of the main text, we have $(p_1, p_2, p_3, q_4)$=(+1, -1, -1,+1) and $(q_1, q_2, q_3, q_4)$=(-1, +1, -1,+1).

The resonance frequencies of each orientation of NV center $(f_\pm)_i$ ($i = 1,2,3,4$) are given by

$$(f_\pm)_i = D + M_{Z_i} \pm \sqrt{(\gamma B_Z)^2 + (M_{X_i})^2 + (M_{Y_i})^2} \quad (1-5).$$

Here, the resonance frequencies $(f_\pm)_i$ ($i = 1,2,3,4$) were determined using the Ramsey sequence described in section 2 below.

We defined $S_i$ as

$$S_i = \frac{f_{+i} + f_{-i}}{2} = D + M_{Z_i} \quad (1-6).$$

We determined the three shear stress components ($\sigma_{xy}$, $\sigma_{yz}$, $\sigma_{zx}$) and the trace of stress tenor ($\sigma_{xx} + \sigma_{yy} + \sigma_{zz}$) by using equations (1-2) - (1-6) as follows:

$$\sigma_{xy} = \frac{S_1 - S_2 - S_3 + S_4}{8a_2} \quad (1-7),$$

$$\sigma_{yz} = \frac{S_1 + S_2 - S_3 - S_4}{8a_2} \quad (1-8),$$

$$\sigma_{zx} = \frac{S_1 - S_2 + S_3 - S_4}{8a_2} \quad (1-9),$$

$$\sigma_{xx} + \sigma_{yy} + \sigma_{zz} = \frac{\frac{S_1 + S_2 + S_3 + S_4}{4} - D}{a_1} \quad (1-10).$$

## 2. Images of the resonance frequencies, $f_{\pm i}$ ($i = 1,2,3,4$) around the dislocation

The resonance frequencies of the NV centers $f_{\pm i}$ ($i = 1,2,3,4$) at each measurement position were measured using a Ramsey sequence. We applied the microwave ($f_{mw}$) to obtain the Ramsey fringes. Figures S2 (b) and (c) show the Ramsey sequence used in this study and Ramsey fringes of $f_{-4}$ measured at position P described in Fig. 3 (b) of the main text. The Ramsey fringes were fitted using the following function,

$$\exp\left(-\frac{\tau}{T_2^*}\right)[a_1\cos(2\pi(f_0-f_h)\tau+\phi_1)+a_2\cos(2\pi f_0\tau+\phi_2)+a_3\cos(2\pi(f_0+f_h)\tau+\phi_3)]$$

where $\tau$, $T_2^*$, $f_0$, $f_h$, $a_i$ and $\phi_j$ ($j = 1,2,3$) are the evolution time, spin dephasing time, microwave detuning, hyperfine splitting, amplitude, and phase, respectively. In this case, the resonance frequency $f_{-4}$ of the NV center is expressed as $f_{-4} = f_{mw} + f_0$. The measurement error of $f_{-4}$, ($f_{-4\text{error}}$) was the same as the fitting error of $f_0$, ($f_{0\text{error}}$). This procedure was performed for the eight resonance frequencies $f_{\pm i}$ ($i = 1,2,3,4$). Figures S3 (a) and (b) respectively show images of the resonance frequencies, $f_{\pm i}$ ($i = 1,2,3,4$), and the measurement error of the resonance frequencies, $f_{\pm i\_\text{error}}$ ($i = 1,2,3,4$), around the dislocation.

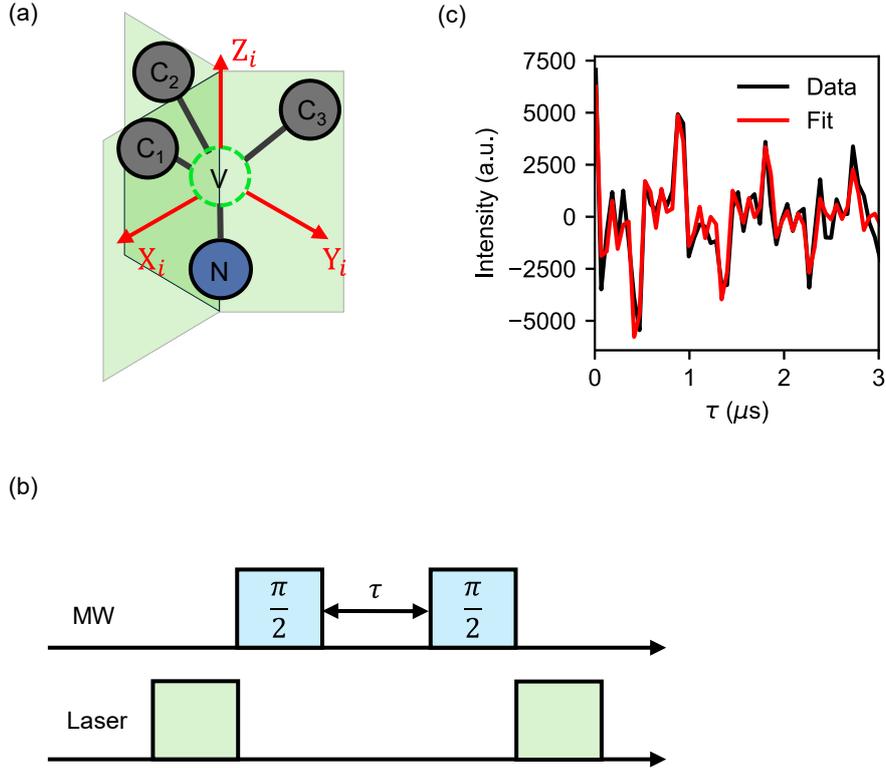

Fig.S2 (a) Schematic of the nitrogen-vacancy center and the adopted coordinate system ($X_i$, $Y_i$, $Z_i$). (b) Ramsey sequence used in this study. (c) Ramsey fringe of $f_{-4}$ measured at position P described in Fig. 3 (b) of the main text. Microwaves with frequencies, $f_{mw}$, of 2709 MHz were applied. The fitting gives $f_0$ = 3.26 MHz with a fitting error $f_{0\text{error}}$ = 0.01 MHz. Thus, the resonance frequency, $f_{-4}$, was estimated to be $f_{-4} = f_{mw} + f_0 = 2709 + 3.26 = 2712.26$ MHz. The measurement error of $f_{-4}$ ($f_{-4\_\text{error}}$) was the same as $f_{0\text{error}}$ of 0.01 MHz.

(a)

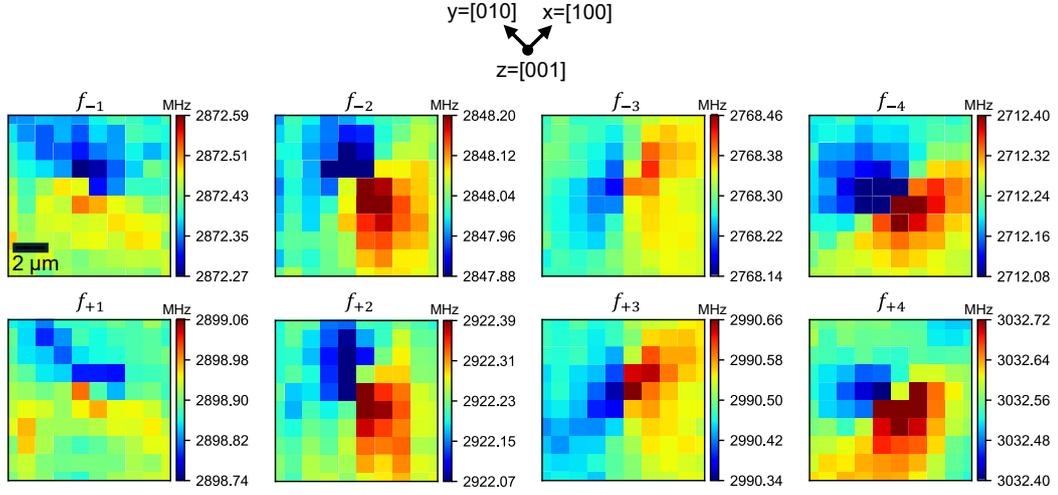

(b)

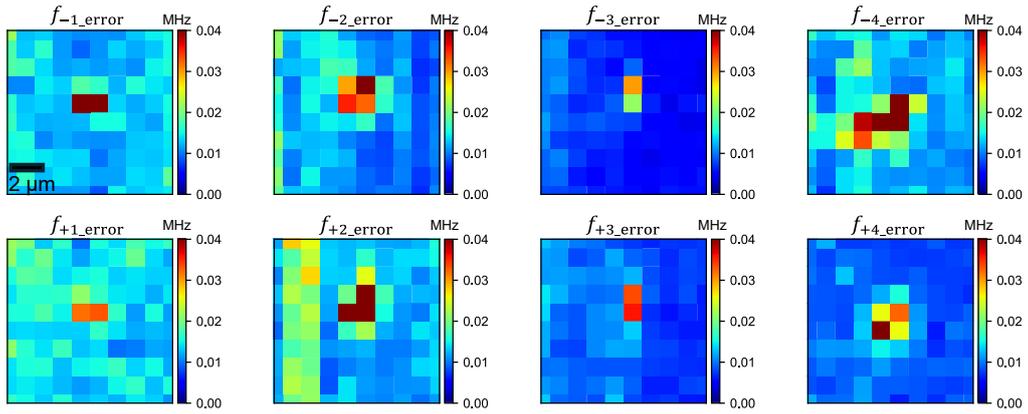

Fig. S3 (a) and (b) Images of the resonance frequencies, $f_{\pm i}$ ($i = 1,2,3,4$) and measurement error of the resonance frequencies, $f_{\pm i\_\text{error}}$ ($i = 1,2,3,4$) around the dislocation, respectively.

## 3. Measurement error of the stress tensor

The measurement errors of the components of the stress tensor, $\sigma_{xy}$, $\sigma_{yz}$, $\sigma_{zx}$ and $\sigma_{xx} + \sigma_{yy} + \sigma_{zz}$ are denoted as $\sigma_{xy\_\text{error}}$, $\sigma_{yz\_\text{error}}$, $\sigma_{zx\_\text{error}}$ and $(\sigma_{xx} + \sigma_{yy} + \sigma_{zz})_{\_\text{error}}$, respectively. $\sigma_{xy\_\text{error}}$, $\sigma_{yz\_\text{error}}$, $\sigma_{zx\_\text{error}}$ and $(\sigma_{xx} + \sigma_{yy} + \sigma_{zz})_{\_\text{error}}$ were contributed to the fitting error of the resonance frequencies $(f_\pm)_i$ ($i = 1, 2, 3, 4$) in accordance with the equations,

$$S_{i_\text{error}}(i = 1,2,3,4) = \frac{\sqrt{(f_{+i_\text{error}})^2 + (f_{-i_\text{error}})^2}}{2} \qquad (3-1),$$

$$\sigma_{xy\text{error}} = \sigma_{yz\text{error}} = \sigma_{zx\text{error}} = \frac{\sqrt{\sum_{i=0}^{4}(S_{i_{\text{error}}})^2}}{8a_2} \quad (3-2),$$

$$(\sigma_{xx} + \sigma_{yy} + \sigma_{zz})_{\_\text{error}} = \frac{\sqrt{\frac{\sum_{i=0}^{4}(S_{i_{\text{error}}})^2}{4}}}{a_1} \quad (3-3),$$

where $f_{\pm i\_\text{error}}$ indicates the fitting errors of the resonance frequencies $(f_\pm)_i$ shown in Fig. S3 (b). Figure S4 shows the images of the measurement errors of the stress tensor around the dislocation. $(\sigma_{xx} + \sigma_{yy} + \sigma_{zz})_{\_\text{error}}$ and $\sigma_{xy\_\text{error}}$ ($= \sigma_{yz\_\text{error}} = \sigma_{zx\_\text{error}}$) were under approximately 0.001 GPa in most areas, while they were over 0.002 GPa near the center of the dislocation.

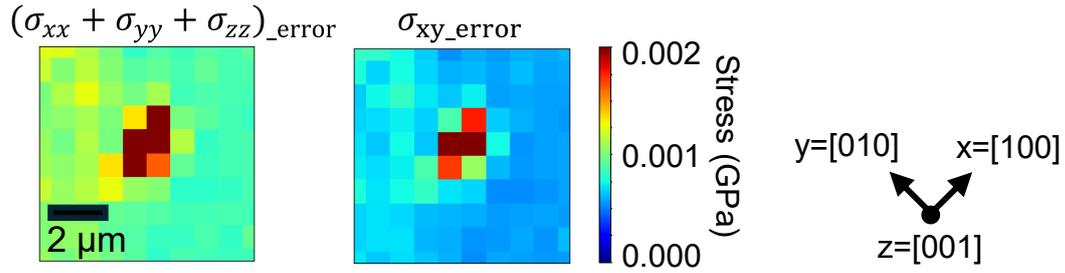

Fig. S4 Images of the measurement errors of the stress tensor, $(\sigma_{xx} + \sigma_{yy} + \sigma_{zz})_{\_\text{error}}$ and $\sigma_{xy\_\text{error}}$ ($= \sigma_{yz\_\text{error}} = \sigma_{zx\_\text{error}}$) around the dislocation.

### 4. Stress tensor around a single edge and screw dislocation

Here we derive the analytical form of stress tensor around single 45°dislocation using elastic theory[3]. As shown in Fig. S5(a), when the dislocation core is located at the origin (0,0,0), the dislocation line vector is along the z=[001] direction, and the Burgers vector of the edge dislocation is along in the x=[100] direction in the x=[100], y = [010] z=[001] coordinate system, stress tensor around single edge dislocation based on the elastic theory [3] are presented as

$$\sigma_{xx} + \sigma_{yy} + \sigma_{zz} = \frac{\mu b_{edge}}{2\pi(1-\nu)} \frac{x^2 y + y^3 + \nu y(x^2 + y^2)}{(x^2 + y^2)^2} \quad (4-1),$$

$$\sigma_{xy} = \frac{\mu b_{edge}}{2\pi(1-\nu)} \frac{x(x^2 - y^2)}{(x^2 + y^2)^2} \quad (4-2),$$

$$\sigma_{yz} = 0 \quad (4-3),$$

$$\sigma_{zx} = 0 \qquad (4-4),$$

where $\mu$ = 578 (GPa) is lame constant and $v$=0.05 is poisson's ratio, $b_{edge}$ is the perpendicular component of the Burgers vector with respect to the dislocation line vector. Please note that while compressive stress and tensile stress were defined as positive and negative, respectively in this study, the elastic theory [3] defines the compressive stress and tensile stress as negative and positive, respectively. Thus, sign (positive/negative) of $\sigma_{xx} + \sigma_{yy} + \sigma_{zz}$ is reversed in the equation (4-1).

Similarly, as shown in Fig. S5(b), when Burgers vector of the screw dislocation is along in the z=[001] direction, stress tensor around single screw dislocation in diamond [3] are presented as

$$\sigma_{xx} + \sigma_{yy} + \sigma_{zz} = 0 \qquad (4-5),$$

$$\sigma_{xy} = 0 \qquad (4-6),$$

$$\sigma_{yz} = -\frac{\mu b_{screw}}{2\pi} \frac{y}{x^2 + y^2} \qquad (4-7),$$

$$\sigma_{zx} = \frac{\mu b_{screw}}{2\pi} \frac{x}{x^2 + y^2} \qquad (4-8),$$

where $b_{screw}$ is the parallel component of the Burgers vector with respect to the dislocation line vector. As mentioned in the main text, since the Burgers vector of the dislocation measured in this study is along the [$\bar{1}$01] direction. Thus, as shown in Fig. S5(c), $b_{edge}$ and $b_{screw}$ are calculated as follows,

$$b_{edge} = -b\cos(45°) \qquad (4-9),$$

$$b_{screw} = b\cos(45°) \qquad (4-10),$$

where the magnitude of the Burgers vector $b = \left|\frac{a}{2}[10\bar{1}]\right| = \frac{\sqrt{2}}{2}a$ ($a$=0.356 nm) of diamond [4]. Elasticity theory has been assumed that the stress tensor components ($\sigma_{xy}$, $\sigma_{yz}$, $\sigma_{zx}$ and $\sigma_{xx} + \sigma_{yy} + \sigma_{zz}$) around single 45° dislocation can be represented as a linear combination of the stress tensors of edge and screw dislocation [3]. Therefore, the stress tensor around a 45° dislocation is expressed as follows using equations (4-1) - (4-10),

$$\sigma_{xx} + \sigma_{yy} + \sigma_{zz} = -\frac{\mu b}{2\pi(1-v)} \frac{x^2 y + x^3 + vy(x^2 + y^2)}{(x^2 + y^2)^2} \cos(45°) \qquad (4-11),$$

$$\sigma_{xy} = -\frac{\mu b}{2\pi(1-v)} \frac{x(x^2 - y^2)}{(x^2 + y^2)^2} \cos(45°) \qquad (4-12),$$

$$\sigma_{yz} = -\frac{\mu b}{2\pi} \frac{y}{x^2 + y^2} \cos(45°) \qquad (4-13),$$

$$\sigma_{zx} = \frac{\mu b}{2\pi} \frac{x}{x^2 + y^2} \cos(45°) \qquad (4-14),$$

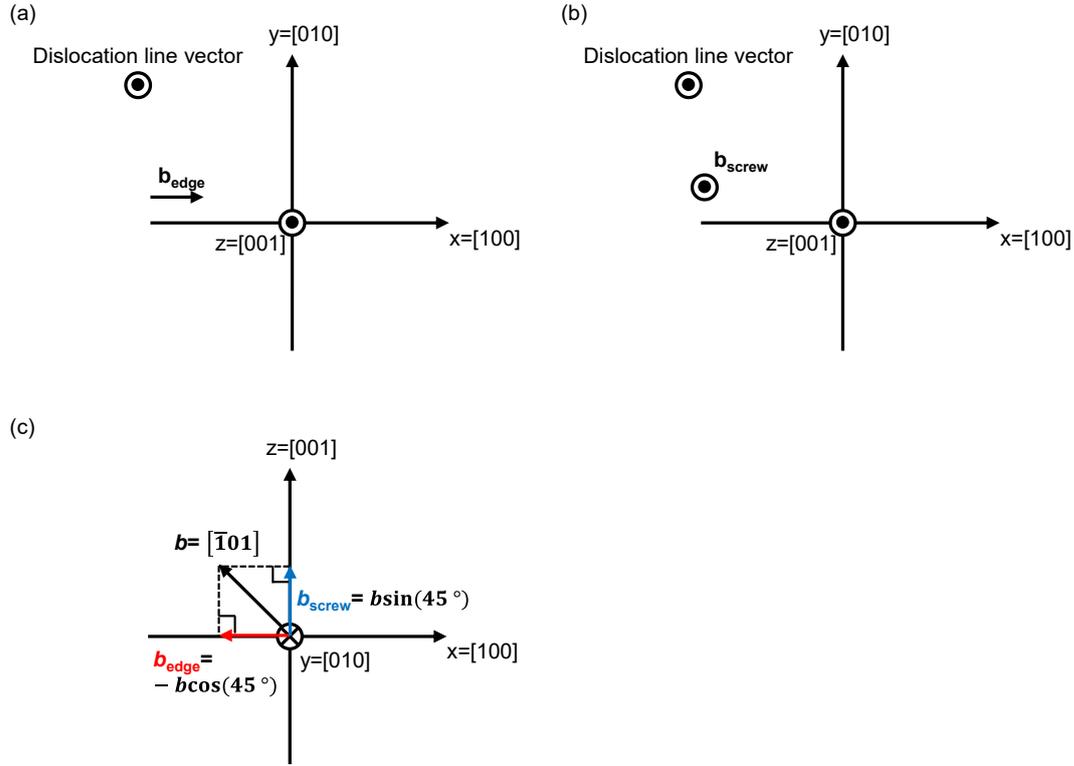

Fig. S5 (a)-(b) Definitions of the dislocation line vector and Burgers vector assumed when calculating the stress tensors around edge and screw dislocations, respectively. (c) A schematic showing the relationship between the Burgers vectors of edge and screw dislocations.

## 5. Comparison of Experimental Data with Elastic Theory

We evaluated the consistency between the experimental results and elastic theory. Specifically, we compared the values of the Lamé constant ($\mu$) and Poisson's ratio ($\nu$) obtained by fitting the experimental stress tensor data shown in Fig.4 of main text to equations (9)–(12) of the main text with the values of $\mu$ and $\nu$ calculated from moduli and stiffness data in previous studies. The fitting parameters of $\mu$ and $\nu$ were obtained by making least-squares fittings to these equations (9)–(12) for each measurement position. In the fitting process, it was not possible to independently determine $\mu$ and the magnitude of the Burgers vector, $b$, because these parameters appear in the equations as a product, $\mu b$. In previous studies, the magnitude of $b$ has been reported to obey $b = \left|\frac{a}{2}[10\bar{1}]\right| = \frac{\sqrt{2}}{2}a$, where $a$ is the lattice constant of diamond [4,5], while $a$ has been reported to be between 0.35666 and 0.35672 nm[6], which means the magnitude of t $b$ would range from 0.252197 nm to 0.252239 nm. Therefore, in our analysis, we fixed the magnitude of the

Burgers vector to $b=0.252197$ nm and treated the Lamé constant $\mu$ as a fitting parameter. As a result, we obtained $\mu = 504 \pm 17$ and $v = 0.036 \pm 0.033$, respectively. The reason for the large error of Poisson's ratio is that, as will be discussed later, the value of $v$ is more than ten times smaller than 1, making $1-v \approx 1$ shown in the equations (9) and (10) in the main text.

Next, we calculated the Lamé constant $\mu$ and Poisson's ratio $v$ based on the Young's module and elastic stiffness values reported in previous studies. Here, $\mu$ and $v$ for an isotropic body such as diamond crystal are expressed as[3]

$$\mu = \frac{E}{2(1+v)} \quad (5-1),$$

$$v = \frac{C_{11}}{C_{12} + C_{11}} \quad (5-2)$$

where E and $C_{11}$ ($C_{12}$) are Young's module and the elastic stiffness constant, respectively. Table S1 shows Young's modulus E, elastic stiffness constants $C_{11}$ ($C_{12}$), Lamé constants $\mu$, and Poisson's ratios $v$ reported in previous studies. On the basis of the data in the previous studies, $\mu$ was 477 [7], 475 [8], 533 [9], and 523 [10] (Gpa) and $v$ 0.10 [7], 0.10 [8], 0.07 [9], and 0.057 [10]. These values are comparable to the results obtained from the fitting, $\mu = 504 \pm 17$ and $v = 0.036 \pm 0.033$. Figures S6 (a), (b), and (c) show the stress tensor measured with the NV center, the theoretical results using the Lamé constant and Poisson's ratio obtained from the fitting, and the theoretical results using the values of $\mu$ and $v$ calculated from moduli and stiffness data in the previous, respectively.

Figure S7 shows the experiment data and the theoretical results based on the previous studies along the diagonal indicated by the black dashed line in Fig. S6. The error bars represent the measurement error of the stress tensor shown in Fig. S4. In the vicinity of the dislocation core (X = 0 µm), the measurement error exceeded 0.01 GPa. This was attributed to the reduced stress sensitivity of the NV center, caused by the high stress variation near the core. In contrast, in regions farther from the dislocation core, the experimental results were consistent with theoretical values. The deviation of each component ($\sigma_{xx} + \sigma_{yy} + \sigma_{zz}$ and $\sigma_{xy}, \sigma_{yz}, \sigma_{zx}$) between the stress tensor measured with the NV center shown in Fig. S5 (a) and the theoretical results based on the previous studies shown in Fig. S5 (c) were 0.013, 0.008, 0.012, and 0.006 GPa, respectively.

| Reference | E (GPa) | $C_{11}$ (GPa) | $C_{12}$ (GPa) | $\mu$ (GPa) | $v$ (GPa) |
|---|---|---|---|---|---|
| [7] | 1054 | 10.79 | 5.78 | 477.76* | 0.103* |
| [8] | 1050 | 10.78 | 5.774 | 475.52* | 0.104* |
| [9] | 1141 | no data | no data | 533.12 | 0.07 |

| [10] | 1106 | no data | no data | 523.23 | 0.0569 |

*$\mu$ and $v$ were estimated using equations (S1) and (S2).

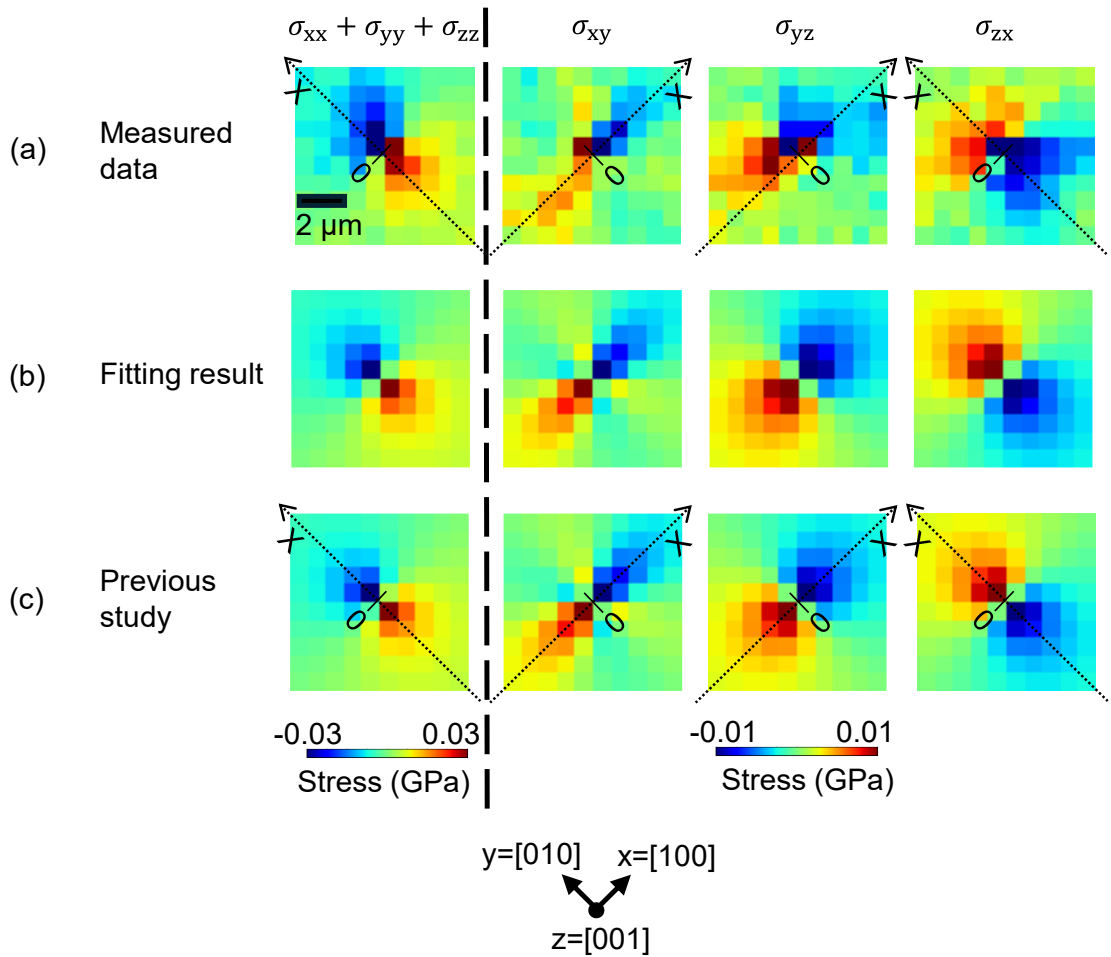

Fig. S6 (a), (b), and (c) the stress tensor measured with the NV center, theoretical results using the Lamé constant and Poisson's ratio obtained from the fitting, and theoretical results using the Lamé constant and Poisson's ratio based on data from previous studies.

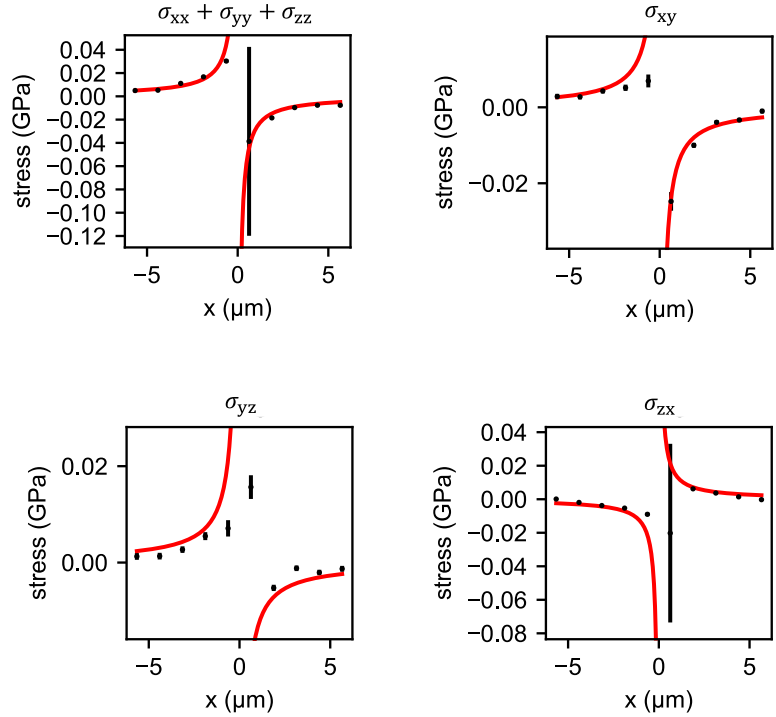

Fig. S7 Experimental data (black dots) and theoretical results using the Lamé constant and Poisson's ratio calculated from data of previous studies (red line) along the diagonal indicated by the black dashed line shown in Fig. S5. Error bars represent the measurement error of the stress shown in Fig. S4.

## 6. Identifying the position of the dislocation in the confocal microscope setup

To identify the position of the dislocation in the confocal microscope setup, we used the X-ray topography images, birefringence microscopy images, and optical microscopy images. First, we identified the positions of the dislocations on the CVD diamond film by using the X-ray topography shown in Fig. S8(a). Here, the position of the dislocation measured in this study is circled in red. Next, we acquired a birefringence image (Fig. S8(b)). The petal shape pattern located on the left side of this image represents the dislocation measured in this study. Next, using the same transmitted light employed in the birefringence image, we obtained the transmitted optical microscope image (Fig. S8(c)). Then, we acquired a reflection optical microscope image (Fig. S8(d)) after placing the copper wire for microwave irradiation on the CVD diamond film. Finally, we searched for the position of the dislocation indicated by the red circle by using confocal microscopy (Fig. S8 (e)) with the location of the copper wire for microwave irradiation used as a reference.

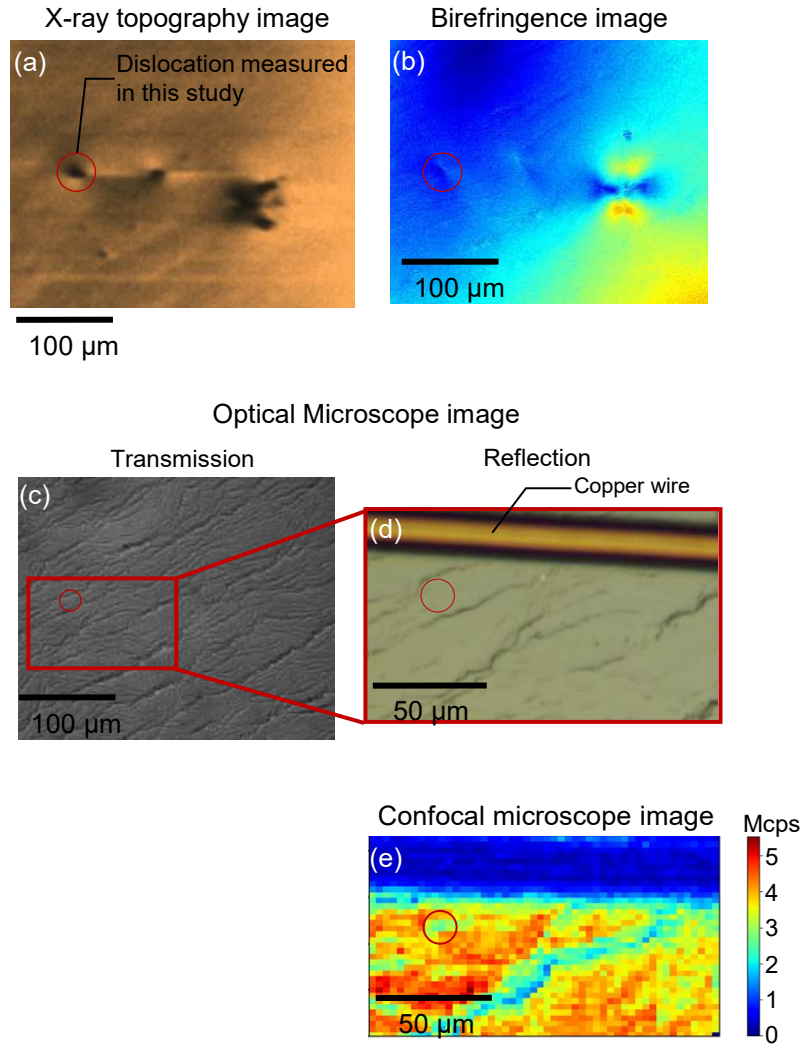

Fig. S8 (a) X-ray topography image of the CVD diamond. The position of the dislocation measured in this study is circled in red. (b) Birefringence microscopy image of the CVD diamond. The petal shape pattern circled in red represents the dislocation measured in this study. (c) Transmission optical microscope image taken in the region shown in Fig. S8(b). (d) Reflection optical microscope image taken in the same region shown in Fig. S8(c). This image was taken after placing the copper wire for microwave irradiation on the CVD diamond film. (e) Confocal microscope image of the CVD diamond film taken in the same region shown in Fig. S8(d). The red circle shows the position of the dislocation measured in this study.